\input{style/arxiv-general.cfg}
\documentclass[aoas,MSNbibl,nameyear,seceqn,dvips]{arximspdf}
\makeatletter
   \@ifpackageloaded{graphicx}{}{\usepackage{graphicx}}
\makeatother

\doi{10.1214/14-AOAS798}% Updated by VTEXPTS2LaTeX.exe, 18.12.2014 08:43
\volume{9}
\issue{1}
\pubyear{2015}
\firstpage{402}
\lastpage{428}
\docsubty{FLA}


\begin{document}
\begin{frontmatter}

\title{Bayesian nonparametric cross-study validation of prediction methods}
\runtitle{Cross-study validation}

\begin{aug}
% Corresponding author: Lorenzo Trippa - ltrippa@jimmy.harvard.edu% Updated by VTEXPTS2LaTeX.exe, 19.12.2014 12:16
%Updated by VTEXPTS2LaTeX.exe, 18.12.2014 08:43
\author[A]{\fnms{Lorenzo}~\snm{Trippa}\corref{}\thanksref{m1,m2}\ead[label=e1]{ltrippa@jimmy.harvard.edu}},
\author[B]{\fnms{Levi}~\snm{Waldron}\thanksref{m3}\ead[label=e2]{levi.waldron@hunter.cuny.edu}},
\author[C]{\fnms{Curtis}~\snm{Huttenhower}\thanksref{m1}\ead[label=e3]{chuttenh@hsph.harvard.edu}}
\and
\author[A]{\fnms{Giovanni}~\snm{Parmigiani}\thanksref{m1,m2}\ead[label=e4]{gp@jimmy.harvard.edu}}
\runauthor{Trippa, Waldron,  Huttenhower and Parmigiani}
\affiliation{Harvard School of Public Health\thanksmark{m1},
Dana-Farber Cancer Institute\thanksmark{m2}
and School of Urban Public Health at Hunter College, City University of New York\thanksmark{m3}}
\address[A]{L. Trippa\\
G. Parmigiani\\
Biostatistics and Computational Biology\\
Dana-Farber Cancer Institute\\
450 Brookline Ave.\\
Boston, Massachusetts 02115\\
USA\\
\printead{e1}\\
\phantom{E-mail:\ }\printead*{e4}}
\address[B]{L. Waldron\\
CUNY School of Public Health\\
\quad at Hunter College\\
2180 3rd Ave Room, 538\\
New York, New York 10035\\
USA\\
\printead{e2}}

\address[C]{C. Huttenhower\\
Biostatistics Department\\
Harvard School of Public Health\\
Building 1 \#413\\
655 Huntington Avenue\\
Boston, Massachusetts 02115\\
USA\\
\printead{e3}}
%
%Biostatistics and Computational Biology\\
%Dana-Farber Cancer Institute\\
%450 Brookline Ave.\\
%Boston, Massachusetts 02115}

%and\\
%Dana-Farber Cancer Institute\\
\end{aug}

% HISTORY:
%
\received{\smonth{12} \syear{2013}}% Updated by VTEXPTS2LaTeX.exe,
%18.12.2014 08:43
%
\revised{\smonth{9} \syear{2014}}% Updated by VTEXPTS2LaTeX.exe,
%18.12.2014 08:43

% ABSTRACT
\begin{abstract}
We consider comparisons of statistical learning algorithms using
multiple data sets, via leave-one-in cross-study validation: each of
the algorithms is trained on one data set; the resulting model is then
validated on each remaining data set. This poses two statistical
challenges that need to be addressed simultaneously. The first is the
assessment of study heterogeneity, with the aim of identifying a subset
of studies within which algorithm comparisons can be reliably carried
out. The second is the comparison of algorithms using the ensemble of
data sets. We address both problems by integrating clustering and model
comparison. We formulate a Bayesian model for the array of
cross-study validation statistics, which defines clusters of studies
with similar properties and provides the basis for meaningful
algorithm comparison in the presence of study heterogeneity. We
illustrate our approach through simulations involving studies with
varying severity of systematic errors, and in the context of medical
prognosis for
patients diagnosed with cancer, using high-throughput
measurements of the transcriptional activity of the tumor's genes.
\end{abstract}

% KEYWORDS
\begin{keyword}
\kwd{Reproducibility}
\kwd{validation analysis}
\kwd{meta-analysis}
\kwd{random partitions}
\kwd{Bayesian nonparametrics}
\kwd{cancer signatures}
\end{keyword}
\end{frontmatter}

\section{Introduction}

Predictive models, in most cases, need to be validated using data from
independent
studies. In many disciplines it is common for research communities to
generate multiple
data sets that address similar prediction problems.
The availability of multiple
data sets makes it possible to systematically compare the performance of
alternative statistical learning algorithms, and to characterize their
strengths and limitations
in the context of a specific area of
application.
% We propose such a systematic approach based on a comprehensive
%analysis of all possible combinations of training and validation
%datasets.

%It also poses currently unaddressed inferential problems,
%which are the subject of this paper.

Here, the term learning
algorithm is used for any procedure, say, linear regression or nearest
neighbor classification, that produces prediction rules.
We consider the task of assessing learning
algorithms, via what we call leave-one-in cross-study validation:
the algorithm is trained on one data set; the resulting prediction
model is then validated on each remaining data set, and a validation
performance statistic (such as the classification error rate or the
mean squared error of prediction) is recorded. By repeating this over
all possible training data sets one generates a square array $Z$ of validation
statistics.
%Unlike available methods for ranking algorithms by cross-validation in
%each of multiple studies,
%his method explicitly addresses the potential for dataset-specific
%bias and the need to evaluate predictions by independent laboratories.
Computation of leave-one-in
matrices $Z$ is, in most cases, straightforward. %, but
%not new in machine learning \citep{}, \\
%{\tt get ref from Levi or Curtis} \\
%statistical
%approaches for the analysis of validation statistics and for comparing
%algorithms in this setting are lacking.
Our
goal is to develop a statistical
framework for the analysis of leave-one-in matrices.

Our motivation
comes from earlier experience in clinical genomics
[\citet{Garrett-Mayer2008}] where the goal is to predict
individual outcomes based on high-dimensional features of the genome.
%%
%measured by high-throughput technologies \citep{Willard2009vr}, and
%where reproducing proposed signatures is an area of significant interest
%%
%%
Leave-one-in cross-study
validation is well suited to this context for two reasons. First, while
different studies address the same prediction question, they may do so
using different sampling designs or technological platforms, generating
heterogeneity that makes it difficult to directly combine all data.
Second, it is not uncommon for studies to be affected by unknown
artifactual variation, such as the so-called batch
effects, %(Leek et al., 2010)
% \citep{leektackling2010}
making it important to use methodologies that
allow identification and separate handling of studies that show poor
concordance with the majority of the rest [\citet{baggerly2008run}].

Our perspective is therefore that cross-study validation
should simultaneously be concerned about two questions: the
identification of heterogeneity and outliers among studies, and the
comparison of alternative algorithms, done in a way that accounts for
heterogeneity across studies. We achieve this by modeling directly
each of the
algorithm-specific $Z$ matrices. Variability in the validation measures
contained in a $Z$ matrix may arise from several
sources, including differences in study design, study populations and
measurement technologies, as well as accidental causes that may have
affected data quality in individual studies.
%and carry out
%separate comparisons of algorithms in each cluster.
To illustrate,
imagine the outcome of interest is determined by a different set of
predictors in different geographical areas. A collection of studies may
include two major clusters of studies, each confined to a given
area. Performance evaluations are best handled by considering cross-study
validation within each of these clusters, as a good algorithm
should not be required to generate models that predict well across
geographical areas when trained on data from a single area. Similar
considerations apply to clusters defined by technological platforms.
%The $Z$ matrix provides information on this
%aspect as well, because such studies will typically provide worse
%cross-study validation performance both when used in training and when
%used in validation.

We propose a two-stage
procedure.
The first stage addresses sampling variation in the $Z$ array via
Bootstrap.
The second stage infers a latent partition of the
studies defined by a Dirichlet process. Studies will be
assigned to the same subset when the corresponding vectors of
validation statistics are
similar. Conversely, if the $Z$ array provides evidence of
heterogeneity between two
studies, then these will tend to be assigned to separate clusters.
%Outlying studies tend to form singletons in the partition.
Our model
achieves two goals: (i) to cluster studies using $Z$, generating
hypotheses on the sources of heterogeneity; and (ii) to provide
cluster-based summaries of algorithm performance, allowing for
comparisons that account for heterogeneity and possible systematic
artifacts in the study pool.
%%
%accounting for uncertainty in cluster assignment, which is generally
%substantial in these applications.
%%
%Direct modeling of $Z$ is
%parsimonious, computationally efficient, and defines clusters in a way
%that is driven by the aspects of study heterogeneity that are directly
%relevant for validation.

% the challenges becomes both estimating the unknown study-specific
%covariates distributions and, importantly,
%In some cases it investigating possible discrepancies across
%study-specific distributions of relevant unmeasured biomarkers.
%%%%%%%%%%%% Introduce the idea of using a meaningful partition (to be
%estimated) of the datasets %%%%%%%%%%%%%

%One of the advantages of modeling the $Z$ array is the possibility of
%estimating, for any pair of studies, the value that the corresponding
%elements of $Z$ would take should both studies be performed a second
%time. Estimates can be derived using the hypothesis that data are
%newly generated under identical technical conditions and that the
%populations from which samples arise remain identical. When the
%estimates are combined with clustering of the studies, these
%contribute to interpretation of the observed values in our $ Z$ array.
%
%It is also important to assess whether a study generated data that is
%sufficiently inconsistent with that of other studies to justify
%exclusion from that analysis of predictions.

Clustering based on the $Z$ matrix is perhaps most attractive in the
context of prediction problems with a large number of predictors. High
dimensionality makes it difficult to spot the important differences
between studies and to understand the factors hindering cross-study
replicability. In this scenario, it is important to provide a solid
evaluation of prediction strategies\vadjust{\goodbreak} using distinct training and
validation data sets. This evaluation should be rooted in the context
of a specific application. The $Z$ matrix helps in this: its strengths
and limitations arise from reducing the problem to a single figure of merit
for prediction performance. It is simple to interpret and easy to
visualize. Also, it is not affected by subtle issues such as
over-fitting, batch effects and selection of favorable training/testing
combinations. The goal of our Bayesian procedure is to retain these
advantages of the $Z$ matrix, to provide an accurate uncertainty analysis
and to suggest clusters for further inquiry.

While the motivation and examples for our methodology come from clinical
genomics, the only requirement for its application is the %public
availability of independent studies using similar approaches to measure
predictors.
%This is an increasingly common situation, and we are therefore
%confident that our methods may be of interest to the statistical
%learning community at large.

\section{Bayesian cross-study validation analysis}\label{sec2}

\subsection{The leave-one-in validation performance matrix $Z$}

We consider a set of $S$ studies, indexed by $s$ and including $n_s$
samples, indexed by $i$.
For study $s$, we have measurements on outcomes $Y_{s,i}$ and predictors
$X_{s,i}$.
%
%the covariates distributions across studies.
%In principle one might attempt the use of a hierarchical model for the
%datasets $\{1,\ldots,S\}$.
%Instead, we make inferences on latent partitions of the $S$ studies
%using a summary representation of the data, consisting of (i) the $ Z$
%array and
%(ii) a parametric estimate $\hat d$ of the unknown joint distribution
%$d$ of the zero mean random variables
%$ Z_{s,v}-\mathbb{E}_{ P_s,P_v}( Z_{s,v})$, where $ s,v=1,\ldots,S $
%and $ s\neq v$. The expected values $\mathbb{E}_{ P_s,P_v}( Z_{s,v})$
%refer to the true unknown distributions of the data $P_s$ and $P_v$
%within studies $s$ and $v$.
%We will provide description of our model and clustering approach
%assuming identical sample sizes $n_1=\ldots=n_S$ across studies, and
%then remove this assumption.
%
Our focus is the two-dimensional array of validation statistics
$Z=(Z_{s,v}; s,v=1,\ldots,S, s\neq v)$.
We use the term algorithm to refer to a training methodology (such as
CART or ridge regression) and the term model to refer to a specific
prediction rule, resulting from using the algorithm on a training data set.
For a given algorithm,
the statistic $Z_{s,v}$ measures the predictive performance of the
model trained on data set $s$, when validated on a different data set $v$.
Typical definitions of $ Z_{s,v}$ with binary outcomes include the
classification error rate and, if the model generates risk scores for
binary outcomes, the area under the operating characteristic curve (AUC).
Validation statistics for time-to-event outcomes include versions of
the concordance index [\citet{unoc-statistics2011} and references therein].
Our approach is based on the $ Z$ matrix and does not include direct
modeling of the data at the individual level.
This choice is motivated by the goal of obtaining easily interpretable
results with modest computational effort.
%As an alternative, one may attempt joint hierarchical modeling of
%multiple datasets.
% However, these require modeling data in much higher dimensional
%spaces, and most of the additional dimensions are not directly
%relevant tot he validation issue.

In addition to $Z_{s,v}$, with $s\neq v$, one can also consider the
variables $ Z_{s,s}$,
obtained by standard cross-validation, iteratively splitting the data
set into training and validation components.
Here we do not use the variables $ Z_{s,s}$ to avoid summary statistics
that might be inflated by systematic errors or batch effects.

\subsection{Relation to Bayesian meta-analysis}

There are important points of contact, as well as differences, between
our approach and existing ideas in Bayesian meta-analysis.
%As a first approximation, one may think of our problem as a
%matrix-valued extension of meta-analysis, made more complex by the
%fact that the goals of algorithm comparison are different from the
%typical goals of evidence synthesis.

Bayesian modeling allows one to easily account for study heterogeneity.
Several approaches are based on hierarchical models [\citet{Berry1990ue}].
For example,
\citet{warnbayesian2002} consider $S=31$ randomized trials for
assessing the analgesic Ibuprofen. % (dose=400 mg).
The data for each study consist of sample size, number of individuals
randomized to placebo and number of events (pain relief) for each arm.
Treatment assignments $X_{s,i}$ and outcomes $Y_{s,i}$ are binary.
They specify a hierarchical model with latent parameters $\theta_s$
describing success rates in each study and an unknown distribution\vspace*{-1pt} $F$
describing variability in the study specific parameters, that is,
$\theta_s|F   \stackrel{\mathrm{i.i.d.}}{\sim} F$.
The assumption that, conditionally on these parameters, individual
observations within each study are independent completes the model.
%Extensions to vectors of effects from related treatments, as in what
%is now called network meta-analysis, have also been studied

Heterogeneity of study-specific parameters is often better understood
via clustering, as we will propose here.
\citet{berryempirical1979} introduced the idea of using a Dirichlet
prior for $F$.
A practical advantage of the Dirichlet process in this context is the
resulting discreteness of $F$.
This implies that when $(\theta_1,\ldots,\theta_S)$ are sampled either
from the prior or from the posterior, one observes %, with positive
%probability,
clusters of studies: for every pair $(s,v)$ the event $\theta_s=\theta
_v$ has positive probability.
Thus, one obtains \textit{a posteriori} the distribution of a latent
random partition of the studies $\{1,\ldots,S \}$ dictated by ties in
the values of the parameters $(\theta_1,\ldots,\theta_S)$.
While evidence synthesis may average over the distribution of this
partition, cluster analysis can be performed by selecting a single
representative partition.
%We will later outline possible approaches for summarizing the law of a
%random partition.
Model-based clustering and the use of a latent partition
are effective for dealing with questions and hypotheses such as
(i) the response probabilities are the same across studies, (ii)~there
exists a large group of studies sharing identical response
probabilities and (iii) there are studies that should be considered outliers.
%For example, in the study illustrated by Warn et al.
%of the response probabilities across studies, but no single study
%stands out as inconsistent with the others.
%Several authors discussed Bayesian models based on random partitions,
%see for example \citep{petermuellermethod2004} and

%We can now consider covariates $X_s$ such as gene expression or other
%biomarkers.
%It is necessary to account for possible discrepancies in the
%covariates distributions across studies.
%In principle one might attempt the use of a hierarchical model for the
%datasets $\{1,\ldots,S\}$.
%Instead, we make inferences on latent partitions of the $S$ studies
%using a summary representation of the data, consisting of (i) the $ Z$
%array and
%(ii) a parametric estimate $\hat d$ of the unknown joint distribution
%$d$ of the zero mean random variables
%$ Z_{s,v}-\mathbb{E}_{ P_s,P_v}( Z_{s,v})$, where $ s,v=1,\ldots,S $
%and $ s\neq v$. The expected values $\mathbb{E}_{ P_s,P_v}( Z_{s,v})$
%refer to the true unknown distributions of the data $P_s$ and $P_v$
%within studies $s$ and $v$. We first provide a description of our our
%model and clustering approach (section \ref{clu}) assuming identical
%sample sizes $n_1=\ldots=n_S$ across studies, and then remove this assumption.

%%%% minimal description of the methodology %%%%%%%%%%%%%
%It is necessary to account for possible discrepancies in the
%covariates distributions across studies.
%In principle one might attempt the use of a hierarchical model for the
%datasets $\{1,\ldots,S\}$.
%Instead,

\subsection{Two-stage analysis}

%For our analysis of the leave-one-in validation data, we propose
%cluster analysis of the $S$ studies using
Our validation analysis uses a summary of the data, consisting of (i)
the $ Z$ array and
(ii) a parametric estimate $\hat{d}$ of the unknown joint distribution
$d$ of the zero mean random variables
$Z_{s,v}-\zeta_{s,v}$, where $s,v=1,\ldots,S$, $s\neq v$, and $\zeta
_{s,v}$ is the expected value of $Z_{s,v}$.
The expected values $\zeta_{s,v}=\mathbb{E}_{ P_s,P_v}( Z_{s,v})$
%are computing with respect
refer to the true unknown distributions of the data $P_s$ and $P_v$
within studies $s$ and $v$. These are joint distributions including
both predictors and outcomes, and might vary across studies.

Our approach is in two stages. % for the analysis of the array $ Z$.
The first stage estimates the dispersion of the $Z_{s,v}$ random
variables. % $( Z_{s,v}; \,\, s,v=1,\ldots,S,\,\,\, s\neq v)$.
%Throughout the article we assume the joint distribution of $( Z_{s,v};
The second stage is based on a Bayesian model, specified using a
Dirichlet prior and the dispersion of the $Z$'s estimated in the first stage.

%%%%%%%% Why different data quality levels. %%%%%%%%%%%%%
%and $P_{v}$ are identical or substantially similar. If the datasets
%have been collected with systematic errors ranging from negligible to
%severely affecting the study results, then ideally we want studies
%with similar
%gravity of errors to be clustered together. If studies focused on
%different subpopulations, but
%a pair of datasets $(s,v)$ considered the same or similar
%subpopulations, then it is again desirable to visualize $s$ and $v$ in
%the same cluster.
%These are the guidelines that we use for constructing the model.
% On the basis of these guidelines the model is constructed so that a
%parsimonious partition $(C_1,\ldots,C_m)$ of $\{1,\ldots,S\}$ tend to
%be indicated as in agreement with the data, if for any pair of
%subsets, say $(C_1,C_2)$, the validation variables $\{ Z_{s1,s2}; \,\,
%variability across this portion of $ Z$ variables can be mainly
%imputed to sampling variability, i.e. these variables do not provide
%evidence against the hypothesis that the unknown sampling models $
%similar.

We propose a simple hierarchical model for $Z$ that balances (i) the
need, as in any validation study, of easily interpretable
summary statistics that are free of questionable assumptions and (ii)
the goal of detecting clusters of studies and possible outliers.
We chose a prior model for $Z$ with a minimal level of complexity in
order to avoid difficulties in the interpretation of the resulting estimates.
Similar to Bayesian meta-analysis, we use latent parameters for the
unknown means of our $Z$ random variables. The
posterior distribution of these parameters, as discussed in Section~\ref{clu}, allows clustering of the studies.
The goal of the model is to cluster studies with similar data quality,
as well as studies sharing similarities in their designs and implementations.
%Here similarity refers to any aspect of the study that might impact
%the performance of the predictive models.
%Overall, our goal is to define a procedure broadly applicable to
%diverse settings.
We will first provide a description of our model and clustering
approach in Section~\ref{clu}, assuming identical sample sizes
$n_1=\cdots=n_S$ across studies, and then remove this constraint.

One of the advantages of modeling the $Z$ array is the possibility of
estimating, for any pair of studies $(s,v)$, the
distribution of $Z_{s,v}$ %the vmetersalue that the corresponding
%elements of $Z$ would take
should both studies be performed a second time.
%We consider this to be a useful and directly interpretable metric for
%reproducibility of validation results.
Estimates can be derived using the hypothesis that data are newly
generated under identical technical
conditions and that the populations from which samples arise remain
identical. When the estimates of $\zeta_{s,v}$ are combined
with the inferred partition of the studies $\{1,\ldots,S\}$,
these contribute to interpretation of the observed values in our $Z$ array.
\begin{longlist}[{Stage}~2.]
\item[\textit{Stage}~1.] The first stage estimates $d$, with the goal of
obtaining an approximate Bayesian analysis for the observed $Z$ array.
The approximation consists of plugging an estimate of $d$ into the
Bayesian model (Stage 2) to bypass computationally intensive joint
modeling of $S$ data sets. A practical method is the Bootstrap, either
in its frequentist [\citet{efron1979bootstrap}] or
Bayesian [\citet{rubin1981bayesian}] versions.
The resulting distribution is representative of the sampling
variability of the $ Z$ statistics.
The observed variations across the $Z$'s are due to both sampling variability
and also to possible differences across the study-specific
distributions $P_1,\ldots,P_S$.

The only result from the first stage of our procedure that we use in
the analysis of the leave-one-in array
is the estimate $\hat{d}$. Alternative estimators of $d$ could in
principle be used. Here we use the bootstrap because of its broad applicability.
It can be applied to $Z$ matrices generated by a spectrum of training
methods ranging from popular machine learning procedures to
algorithms highly tailored to specific application areas. Also, the
bootstrap can estimate the variability of a number of possible
validation summaries,
such as the misclassification error rate or the mean squared error,
that can be used to define
$Z$ arrays. Finally, the bootstrap is applicable wether or not there
exists a probability model consistent with the training algorithm.

% The modeling strategy that we propose can be used in combination with
%alternative methods one wants to consider for the upfront estimate of
%the unknown distribution $d$.

The Bootstrap [\citet{efron1979bootstrap}] for estimating $d$ includes
(i) the computation of the empirical distributions $\hat{P}_{1},\ldots,\hat{P}_{S}$,
which (ii) are then iteratively used for obtaining $S$ independent
Bootstrap samples, one for each study,
$(X_{1,i}^*,Y_{1,i}^*; i\le n_1),\ldots, (X_{S,i}^*,Y_{S,i}^*; i\le
n_S)$, with $(X_{s,i}^*,Y^*_{s,i})\sim\hat{P}_{s}$.
Here we avoid the use of an additional index enumerating Bootstrap iterations.
At each iteration the validation statistics are computed on the basis
of $(X_{1,i}^*,Y_{1,i}^*; i\le n_1),\ldots, (X_{S,i}^*,Y_{S,i}^*; i\le
n_S)$, that is,
the $ Z$ array is resampled.
At each cycle we compute a prediction model using
$(X_{s,i}^*,Y_{s,i}^*; i\le n_s)$ and then validate\vspace*{1pt} it on
$(X_{v,i}^*,Y_{v,i}^*; i\le n_v)$, $s\neq v$, to obtain $ Z_{s,v}^*$.
Finally, (iii) we estimate $d$ by centering the empirical distribution
of the iteratively resampled arrays. % $( Z_{s,v}^*, s,v=1,\ldots,S,\;s
The Bootstrap estimate of $d$, as the number of iterations increases,
converges to the nonparametric maximum likelihood estimate of $d$.
In other words, by resampling we approximate the mapping of $\hat
P_1,\ldots, \hat P_S$ to the distribution of
$( Z_{s,v}-\zeta_{s,v};  s,v=1,\ldots, S, s\neq v)$ under the
assumption that $ P_s=\hat P_s$ for every $ s\le S$.

When $d$ is estimated by the Bayesian Bootstrap, the flow of the
procedure remains identical, with the exception that the initial
components $\hat P_{1},\ldots, \hat P_{S} $ are replaced by random
distributions $P^*_{1},\ldots, P^*_{S}$. The random distributions
$P^*_{1},\ldots, P^*_{S} $ are defined by $P^*_s\propto\sum_{i\le n_s}
W_{s,i} I_{(X_{s,i}, Y_{s,i})}%/(\sum_{i\le n_s} W_{s,i} )
$,
where\vspace*{1pt} $W_{s,i}$, $i\le n_s$, are independent exponential variables with
a fixed scale parameter.
The Bayesian Bootstrap averages over iteratively generated random
distributions $P^*_{1},\ldots, P^*_{S} $.
In this case, the resampling scheme allows us to obtain
% empirical distribution of the resampled validation arrays converges to
the Bayesian estimate of the $Z$'s dispersion
% the unknown distribution of $Z$,
% $( Z_{s,v} ; \,\, s,v=1,\ldots, S,\,\, s\neq v)$,
under Dirichlet process priors with infinitesimal
concentration parameters for $P_1,\ldots, P_s$.

\item[\textit{Stage}~2.]
We specify a Bayesian model for the validation statistics $Z$.
To simplify posterior\vspace*{1pt} computations, we
plug in a zero mean multivariate normal distribution $\hat{d}$ into our
model by matching the covariance matrix estimate from the Bootstrap
algorithm in the previous paragraph.
%approximate the Bootstrap estimates of $d$ with a zero mean
%multivariate normal distribution $\hat d$ by matching the covariance
%matrix.
This choice, in several cases, is justified by convergence of the
actual joint
distribution of the validation statistics~$Z$, for large sample sizes,
to a Normal density. We will provide examples of such convergence.

%Recall that we plug in the estimate $\hat d$ in the model. % and
%consider it as the true $ Z$ centered distribution.
We introduce an exchangeable random partition $\Pi=\{C_1,\ldots,C_m\}$
of $\{1,\ldots, S\}$, where $C_j$, $j=1,\ldots,m$
are groups of studies.
The number of clusters $m$ is a random variable.
The random partition $\Pi$ of $\{1,\ldots,S\}$ is specified by $S$
exchangeable variables
sampled from a discrete random distribution; the Dirichlet process is
an example.
We refer to \citet{lee2013defining} for an overview on exchangeable partitions.
We use $C(s)$ for indicating the subset of the partition $\Pi$ that
includes study~$s$. Also, we
%If, say $S=4$ and $\Pi=[C_1=\{1,3\},C_2=\{2,4\}]$, then $C(1)=C(3)=1$
%and $C(2)=C(4)=2$.
use $p_{\Pi}$ to denote the law of the random partition.
We state the probability model for $ Z$;
it includes a latent partition and a set of random variables $(\mu
_{i,j}; i,j=1,\ldots)$
which play a role similar to the atom locations in a Dirichlet process mixture:
\begin{eqnarray}
\mu &=& ( \mu_{i,j}; i,j=1,\ldots)\mathop{\sim}^{\mathrm{i.i.d.}}
p_{\mu},
\nonumber
\\
\Pi &\sim & p_{\Pi},
\nonumber
\\[-8pt]
\label{fst}
\\[-8pt]
\nonumber
\varepsilon &=& (\varepsilon_{s,v}; s,v=1,\ldots ,S, s\neq v)
\sim\hat{d} \quad \mbox{and}
\\
Z_{s,v}&=& \mu_{C(s),C(v)}+\varepsilon_{s,v}, \qquad s,v=1,
\ldots,S, s\neq v,
\nonumber
\end{eqnarray}
where % $\varepsilon=(\varepsilon_{s,v}; \,\,\, s,v=1,\ldots,S,\,\,\, s
the components $\mu,\Pi$ and $\varepsilon$ are a priori
independent and $p_{\mu}$ is a distribution on the real line.
%Expression (3) specifies an identity in distribution.

The probability that the conditional expected values of a pair $(s,v)$
of $Z$ columns (or rows) are identical is strictly positive:
%more formally: refer
\[
p \biggl( \bigcap_{r\le S} \{ \mu_{C(s),C(r)}=
\mu_{C(v),C(r)}, \mu_{C(r),C(s)}=\mu _{C(r),C(v)} \} \biggr)>0.
\]
Also, the distribution of the array $( \mu_{C(s),C(v)};
s,v=1,\ldots,S,s\neq v )$ is invariant with respect to any
permutation $\sigma=(\sigma_1,\ldots,\sigma_S)$ of $\{1,\ldots,S\}$,
\[
(\mu_{C(s),C(v)};\,\, s,v=1,\ldots,S)\,\mathop{=}^d \,(
\mu_{C(\sigma_s),C(\sigma_v)}; s,v=1,\ldots,S).
\]

The model can handle an arbitrary number of additional studies
$(S+1, S+2,\ldots)$.
%Expressions (1-3) specify a well defined probability model.
%In principle, the investigator can define all components of the prior
%$(p_{ Z},p_{\Pi},d)$ and compute posterior inference on
%the underling clustering structure $\Pi$ and the latent variables $(
%Z_{C(s),C(v)};\,\,\,\, s,v=1,\ldots,S,\,\,\,s\neq v) $.
Therefore, one can perform predictive inference by considering a future
$(S+1)$th study and obtain, conditionally on the observed $ Z$
statistics, the distribution of $(\mu_{C(S+1),C(s)},\mu_{C(s),C(S+1)}; s=1,\ldots,S)$.
%This is the approach that we follow, with the relevant exception that
%we compute a smooth estimate of the unknown distribution $d$, which is
%then plugged into the probability model.
%Once we plug in this component, we proceed with the Bayesian model (

Arrays with exchangeable rows and columns have been studied in a series
of papers beginning with the contributions
of \citet{davidj.representations1981} and \citet{hooverrow-column1981}.
These authors proved de~Finetti-type representations for these processes.
Random arrays invariant in distribution to any simultaneous permutation
$\sigma$ of rows and columns, such as $(\mu_{C(s),C(v)}; s,v\ge1)$,
are called jointly exchangeable.
This type of arrays arises, for instance, when relationships between
individuals are represented using two-way
tables [\citet{roy2009mondrian}]. %; see, for an example, the
%application of Aldous-Hoover results in \citep{roy2009mondrian}.
%Importantly,
In our study, these representation theorems provide a formal
justification to use latent cluster membership variables for modeling
exchangeable arrays. %In our case the latent variables
%are cluster membership indicators.
%It is indeed stated that, given a generic jointly exchangeable array
%$(W_{s,v};\,\,\, s,v\ge1)$, there exist a random sequence
%$(\epsilon_s; s\ge1)$ and a random parameter $\$ such that the
%conditional distribution $p(W_{s,v};\,\, s,v\le S \mid\epsilon_1,
%can be factorized and rewritten as $p(\mid\epsilon_s,\epsilon_v,
\end{longlist}

\subsection{Asymptotic normality of validation arrays}
The proposed model for $Z$ is closely connected with Dirichlet process
mixtures.
% A simple example helps to clarify this connection.
Consider, for example, $S$
studies designed for estimating $\theta_s=\mathbb{E}(Y_{s,i})$.
% Under this circumstance one might be interested to explore the
%hypothesis
%of multiple clusters defined by studies with identical means.
A possible approach for exploring the hypothesis of multiple clusters
defined by studies with identical means $\theta_s$ consists in
combining approximate likelihood functions
$N(\bar{Y}_s=\sum_i Y_{s,i}/n_s;\theta_s, \hat\sigma_s^2/\sqrt n_s)$
with a random distribution $F$ for the means, that is, $\theta_s| F
 \stackrel{\mathrm{i.i.d.}}{\sim}  F$.
%, i.e. $\theta_s\mid F\buildrel iid \over\sim F $.
See \citet{burrbayesian2005} for a detailed study of this approach,
and \citet{dersimonianmeta-analysis1986} for a frequentist perspective.
The approximation, from a Bayesian standpoint, %that usually elicits a
%joint prior for all involved random quantities,
consists in using Normal kernels with scale parameters $\sqrt{\sum_i
(Y_{s,i}-\bar Y_{s})^2}/n_s$, and is supported by asymptotic arguments.
%This approximation is usually motivated by centra
Similarly, we combine an exchangeable random partition with a
multivariate Normal kernel $\hat{d}$ justified, in several cases, by
asymptotic arguments.

A smooth estimate of $d$ is computationally convenient and circumvents
artifacts that arise with a discrete one, including
the possibility of
posterior distributions assigning exactly null probability to most of
the $\Pi$ configurations.
%Smooth estimates of $d$ can be obtained by assuming Normal shape for
%$d$, with estimated covariance $\hat\Sigma_d$ obtained via Bootstrap
%method, or using alternative
%procedures such as smoothing
%with Normal kernels convolution the Bootstrap estimate $\hat d$, i.e. $
%is a kernel covariance selected for smoothing $\hat d$.
%We use multivariate Normal approximation, which in several cases can
%be suggested by asymptotic arguments.
One can identify several cases in which the leave-one-in array is
asymptotically Normal.
%Readers uninterested to specific examples, involving some
%technicalities, can skip these paragraphs.
Below we briefly discuss one case where $Z$ converges to a multivariate
Normal distribution on a linear subspace of $\mathbb{R}^{S\times(S-1)}$.
We discuss results for logistic regression, Poisson regression,
proportional hazards models and support vector machine procedures in
the supplementary material [\citet{supp}].
%The validation metrics used in these examples have been studied in the
%literature
%
%We mention that, if the estimate $\hat\Sigma_d$ is nearly singular,
%then it can be replaced with $\hat\Sigma_d+I \epsilon$, where $I$ is
%the identity matrix and $\epsilon$ is a small positive value, without
%alteration of the overall interpretation of the procedure.
%For each example we outline the proof of asymptotic normality.
% indicate the main arguments %one can use
%for verifying asymptotic normality.

Consider the linear model $Y_s| X_s\sim N(X_s\beta_s, I\sigma_s^2)$,
with $(Y_s,X_s)=(Y_{s,i},X_{s,i};\break  i\le n_s)$, least squares
estimates $\hat\beta_s$ and
mean squared errors (MSE) of prediction
\[
%&\label{e1}
Z_{s,v} = \frac{ \|Y_v-X_v\hat\beta_s\|^2}{n_v}. %=\! \frac{\big( \!Y_v-X_v\hat\beta_v +X_v (\hat\beta_v-\hat\beta_s)\!
\]
%
%See Diebold and Mariano (1995) for generalizations of this loss
%function.
%Consider
Here, and in all the examples in the Supplementary Material, we let all
sample sizes grow at the same rate, %that is let
$n_s\approx c_s n_1$, $s=2,\ldots,S$,
and fix
$c_2,\ldots,c_S$.
Independence of
$ \|Y_v-X_v\hat\beta_v \|^2$
and
$(X_v,\hat\beta_v)$
implies, under mild assumptions on the $X_{s,i}$ distributions,
asymptotic normality. % for $ Z$. % for our array.
First,
$n_v^{-1/2} (\|Y_v-X_v\hat{\beta}_v \|^2-n_v \sigma_v^2)\rightarrow N(0,
2\sigma_v^2)$.
Next, to obtain $Z_{s,v}$,
we need to add to the in-sample mean
squared error
$n_v^{-1}\|Y_v-X_v\hat\beta_v\|^2$ a second term,
$n_v^{-1}[\| X_v ( \beta_v-\beta_s)\|^2+2( \delta_v-\delta_s)X_vX_v ( \beta_v-\beta_s)+\|X_v ( \delta_v-\delta_s)\|^2]$,
with
$\delta_v=(\hat\beta_v-\beta_v)$.
It can be shown that $n_v^{-1/2}(\delta_v-\delta_s) [X_vX_v
( \beta_v-\beta_s)-\mathbb{E}(X_vX_v (\beta_v-\beta_s)
)]\rightarrow0$
and
$n_v^{-1/2}( \delta_v-\delta_s)X_vX_v ( \delta_v-\delta
_s)\rightarrow0$.
Finally, both
$n_v^{-1/2} (\delta_v-\delta_s) \mathbb{E}(X_vX_v (\beta
_v-\beta_s))$
and
$n_v^{-1/2}( \beta_v-\beta_s)(X_vX_v-\mathbb{E}(X_vX_v)) (
\beta_v-\beta_s) $
converge to normal densities. Asymptotic joint normality for $ Z$
follows from the asymptotic independence of $\delta_v$ and
$n_v^{-1/2}(X_vX_v)$.

\section{Cluster-based validation statistics}\label{clu}

The procedure we propose generates a posterior distribution $p(\Pi|Z)$ for the unknown partition $\Pi$ of our $S$ studies.
The tuning of the distribution $p_{\Pi}$ and approaches for selecting
the prior model are discussed in the supplementary material [\citet{supp}].
A\vspace*{1pt} representative partition summarizes the posterior distribution. %as a
%single partition.
We select an estimate
$\hat{\Pi}$ that minimizes the expectation of a loss function $l(\hat{\Pi},\Pi)$, that is, $\hat{\Pi}=\arg\min  \mathbb{E}(l(\cdot, \Pi)| Z)$.
The partition $\hat{\Pi}$ is a posterior point estimate.
\citet{quintanabayesian2003} give a discussion
on the decision theoretic paradigm applied to random partitions.
Several loss functions $l(\hat{\Pi},\Pi)$ have been proposed; see, for
example, \citet{Denud2006}.

We use the easily interpretable \textit{maximum transfer metric}; see
%%Charon et al.(2006)
\citet{charonmaximum2006} for a recent contribution.
This metric $l(\Pi_1 ,\Pi_2)$ is defined as the minimum number of
elementary corrections
necessary to match the partitions $\Pi_1$ and $\Pi_2$;
an elementary correction consists of moving a unit to a different
(possibly empty) subset. If we consider, for example, $\Pi_1=(\{1,2\},\{
3,4\})$ and $\Pi_2=(\{1,4\},\{2,3\})$, then $l(\Pi_1,\Pi_2)=2$, and a
possible chain of
corrections is $(\{1,2\},\{3,4\})\rightarrow(\{1,2,3\},\{4\}
)\rightarrow(\{1,3\},\{2,4\}) $.

%Uncertainty on the unknown partition $\Pi$ can be visualized.
%It is standard practice, for a real parameter of interest, to report
%intervals accumulating specific levels
%of posterior probability. Analogously, the map $k\rightarrow p( \{\Pi
%: l(\hat\Pi, \Pi)\le k\} | Z)$
%can be used for illustrating posterior uncertainty.
%Another graphical tool for $p(\hat\Pi| Z )$ is a heat map,
%showing for
%pairs of studies $(s,v)$, posterior probabilities of the events $\{
%C(s)=C(v)\}$.

Our procedure tends to assign studies to separate clusters
when they differ on aspects that affect the validation statistics $Z$.
The dissimilarity captured by the clustering method might be due to
different measurement techniques, different predictors distributions or
other factors varying across studies.
% In such cases it is attractive to fit models after combining multiple
%studies into a single dataset.
Interpretation of the inferred partition
requires subsequent analyses to identify the primary causes of
heterogeneity, such as data quality or experimental designs.
%, or different distributions of relevant predictors across studies.
The results can then inform the construction of models trained on
multiple data sets.
If, for instance, heterogeneity is driven by different distributions
of relevant predictors, but the covariates
effects on the outcome are consistent across studies, then
it might be appropriate to combine the available data sets.
In contrast, if heterogeneity is driven by measurement errors or batch effects,
additional efforts may focus on data normalization steps.

We can now introduce the concept of clustering-based validation
performance measure, by which we mean
summary statistics aimed at assessing cross-study prediction taking
into account
study heterogeneity and within-cluster similarities.
Recall that model \ref{fst} formalizes the identity between the
conditional expected values of $Z_{s,v}$ and $Z_{s',v'}$
when study $s$ clusters together with $s'$ and $v$ clusters with $v'$.
%The subset can be a singleton.
For example, we may be interested in the performance measure obtained
when one trains on any of the studies in the cluster of study $s$ and
validates in any of the studies in the cluster of study $v$, that is,
$\mu_{C(s),C(v)}$.
The latent variable $C(s)$ indicates the cluster that includes $s$ and
$ \mu_{C(s),C(v)}$ can be interpreted as the expectation of $ Z_{s,v}$
assuming that studies $s$ and $v$ are repeated de novo.
% under the unknown distributions $P_s$ and $P_v$.
A point estimate $\mathbb{E}(\mu_{C(s),C(v)} | Z)$ can be obtained
by averaging $\mathbb{E}(\mu_{C(s),C(v)} |\Pi, Z)$ with respect to
the posterior distribution of the partition $\Pi$.
Similarly, one may derive interval estimates.
%(levi comments: this next sentence doesn't make sense to me)

We can also estimate the validation performance that one would obtain
from training in a study from the set $C(s)$, and validating in a
future $(S+1)$th study, by using $ \mu_{C(s),C(S+1)}$.
In particular, the joint posterior distribution of $ \mu_{C(s),C(S+1)}$
and $\mu_{C(v),C(S+1)}$, with $s,v\le S$, can be used for comparing
studies $s$ and $v$.
%Finally $ \mu_{C(S+1),C(S+2)}$
%Note that,
%considers two future datasets.
%, by exploiting the
%fact that the random partition $\Pi$ can be extended.
% to an arbitrary number of studies.

Let $B$ be a subset of studies in $ \{1,\ldots, S\}$. We extend the
definition of the validation statistic $Z_{s,v}$ to handle the case
where a model is trained on the combination of the data from all the
studies in $B$, and then validated on study $v$.
We denote the resulting validation statistic by $Z_{B,v}$. If $B$
includes $v$, then $v$ is not used to train the model, and $Z_{B,v}$ is
redefined to be the same as
$Z_{B\setminus v ,v}$, where $B\setminus v =\{s\le S \dvtx s\in B \mbox{ and } s\neq v \}$.
We also use $B(s)\subset\{1,\ldots, S\}$ to denote the studies within
the same $\Pi$ latent cluster of $s$, that is,
$B(s)=\{ v\le S\dvtx  C(s)=C(v) \}$.
%GPQ{\tt what is the difference between B(s) and C(s): aren't they both
%the same cluster? don;t they both live in the power set of 1...S?}

Clustering has the goal of identifying homogeneous groups of studies
with similar sampling distributions. When this works, it is natural to
train models by combining the studies in a cluster.
However, the figure of merit used for the $Z$ summary, not unlike a loss
function, implies adopting a specific one-dimensional perspective in
looking at the data. It is possible, for example, that two studies with
different covariate distributions might be clustered together, or two
studies which only differ in design, but not in the populations, may be
allocated to separate clusters.

Clustering can be used to estimate the performance obtained when
validating in study $s$ after training on studies in
$B(s)$, that is, using only data sets similar to $s$. This task
reduces to estimating $Z_{B(s),s}$.
The function $B\rightarrow Z_{B,s}$, over the collection of $\{1,\ldots
,S\}$ subsets, can be directly computed using our $S$ data sets and is
not related with
the Bayesian model, but $Z_{B(s),s}$, the value of this function at
$B(s)$, is estimated because $B(s)$ is an unknown latent component
of the model.
This approach is only useful when there is no strong evidence that $s$
belongs to a singleton cluster.
%From a Bayesian perspective,
We thus estimate $Z_{B(s),s}$ by using the posterior distribution of
the partition $\Pi$
%%
%Let $M_{B}$, given a generic subset $B\subset\{1,\ldots, S\}$, denote a
%model trained by the merged datasets in $B$, let $ \mathcal{Z}(M, s)$
%be a statistic
%assessing the prediction accuracy of a model $M$ obtained by a
%validation on dataset $s$, and let
%$B(s)\subset\{1,\ldots, S\}$ indicates the studies within the same
%latent cluster of $s$.
%With the introduction of an unknown partition $\Pi$ one can attempt
%to estimate how well the outcomes in $s$ could have been predicted on
%the basis of
%datasets in $B(s)$. Such estimate is relevant when there in no evidence
%in favor of $B(s)=\{s\}$.
%From a Bayesian perspective, the estimate can be derived either using
%the posterior distribution of $\Pi$
%%
%
%it is necessary to
and conditioning on the event $B(s) \neq\{s\}$.
We report both the estimate of $Z_{B(s),s}$ obtained by averaging over
$\Pi$ configurations with $B(s) \neq\{s\}$
and the posterior probability of the conditioning event $B(s) \neq\{s\}$.
%, for instance $p(B(s) \neq\{s\} | Z)$.
Alternatively, we can generate a plug-in estimate $Z_{\hat B(s),s}$ by
focusing on $\hat B(s)$, the cluster in $\hat\Pi$ that includes the
$s$th study.

When we estimate $Z_{B(s),s}$ the goal is to evaluate a model trained
by a homogeneous set of studies $B(s)$.
Our clustering procedure uses validation statistics to detect study
heterogeneity, and therefore the resulting
partition is representative of differences between studies captured by
the $Z$ validation summaries.
Studies included in the same cluster could still differ in important
ways. We consider this point further in the discussion.

% involves
%assessing
% $M_{\hat B(s)\tiny\setminus\normalsize s}$, where $\hat B(s)\tiny
%estimated cluster $\hat B(s)$; i.e. if
%$\hat B(s)\tiny\setminus\normalsize s\neq\emptyset$, then $M_{\hat
%B(s)\tiny\setminus\normalsize s}$ can be validated by $
%The second approach involves conditioning on $B(s)\tiny\setminus
%$ \mathcal{Z}(M_{B(s)\tiny\setminus\normalsize s}, s)$ becomes a
%well defined function of $\Pi$ and, $\mathbb{E}( \mathcal{Z}(M_{B(s)
%trained by $B(s) \tiny\setminus\normalsize s$.

For comparing studies, we also need to be concerned about the potential
for variations
of clustering-based summaries,
such as $Z_{B(s),s}$,
driven
by different total sample sizes within each cluster.
Under the assumption of identical sample sizes $n_1=\cdots=n_S$, which
will be later removed, one can
expect that the value $Z_{B(s),s}$ improves with the number of studies
in $B(s)$.
We thus define the \textit{sample size adjusted} validation statistics
$Z^j_{B,s}$.
%GPQ\\{\tt shouldn't we do this before we even define B? what if a
%study ends up as a singleton just because it is very small?}\\
The definition of these statistics is analogous to that of $Z_{B,s}$.
%While other approaches are possible, we consider the ensemble of
%datasets in cluster $B$ to form a collection of exchangeable samples.
We randomly select $j$ distinct samples from the ensemble of studies~$B$.
We train a model on these $j$ samples and
validate it on data set $s$ to generate a performance measure, say, an AUC.
We iterate this procedure, keeping fixed both $B$ and $s$; $Z^j_{B,s}$
is the average of the accuracy measures obtained
during these iterations.
In this case, if $B$ includes $s$, then the units in $s$ are not
selected for training the model.
The index $j$ can vary from a minimal size of interest up to the
overall number of samples in $B \setminus s$.

Our\vspace*{-1pt} interest is in the map $j\rightarrow Z^j_{B(s),s}$; recall that
$B(s)$ is unknown but can be estimated using
the posterior distribution of $\Pi$. The statistics $Z^j_{B(s),s}$ have\vspace*{-1pt}
an interpretation similar to $Z_{B(s),s}$; moreover, one can contrast
the estimates of $Z^j_{B(s),s}$
and $Z^j_{B(v),v}$ to compare the $s$th study to the $v$th study.
We can estimate $Z^j_{B(s),s}$ plugging in the point estimate $\hat\Pi
$ or directly using\vspace*{1pt} the posterior distribution of~$\Pi$.
If we follow the first approach, the estimator is $Z^j_{\hat B(s),s}$,
while the second approach
%the estimator is $\mathbb{E}( Z^j_{B(s),s} | Z, \sum_{B(s)} n_v \ge
%j+n_s)$, where the
averages with respect to the posterior distribution of $B(s)$.
In both cases we estimate, assuming %the existence of a latent
%partition $\Pi$
%GPQ{\tt why is this conditional on pi?}
$\sum_{ B(s)} n_v \ge j+n_s$, the mean value of the validation
statistic when
the algorithm is trained by $j$ data points from the unknown subset
$B(s)\setminus s$ and then validated on $s$.
In the second case, we report the posterior probability of $ \sum_{B(s)} n_v \ge j+n_s $,
and compute our estimate conditionally on this event
because $Z^j_{B(s),s}$ is well defined only when $B(s)$ includes at
least $j+n_s$\vspace*{-3pt} units.

\section{Simulation study}

\subsection{Scenario 1}\vspace*{-6pt}

The goal of this simulation study is to illustrate the extent to which
our model-based approach contributes to the interpretation of
cross-study validation statistics, beyond what can be learned from
direct visualization of $Z$.
As this relies on estimating the unknown partition $\Pi$ and the
latent $\mu_{C(s),C(v)}$ variables, we also discuss our model's ability
to reconstruct these.

The scenario is defined by 9 studies grouped into three
clusters, $C_1=\{1,2,3\}$, $C_2=\{4,5,6\}$ and $C_3=\{7,8,9\}$, which
differ in the amount of measurement error in the predictors. All
studies have a sample size of 300.
For subject $i$ from study $s$ we have a binary outcome $Y_{s,i}$ and
50 candidate predictor variables $X_{s,i}$.
In group $C_1$, the 50 covariates are simulated from a multivariate
Normal distribution with null mean;
all variances are equal to 17. %, and all covariances are equal to 16.
The dependence between $X_{s,i}$ and $Y_{s,i}$, $s=1,2,3$, is
specified by a logistic regression function; 10 regression coefficients
are equal to 0.1 and 40 are equal to 0.
In group $C_2$ we add independent measurement errors with null mean and
standard deviation equal to 14 to 50\% of the covariates.
In $C_3$ we add independent measurement errors with mean 0.33 and
standard deviation 8 to all covariates. %**a sentence about how to
%interpret C2 and C3.**

For each study we obtain a prediction model by fitting a logistic
function using ridge regression; we tune the penalization parameter
with standard cross-validation.
We then assess model performance using the mean absolute error (MAE)
of prediction, that is, $Z_{s,v}= n_v^{-1}\sum_i \|Y_{v,i}- \mathrm{logit}^{-1} (\hat\beta_s^o+\hat\beta_s X_{v,i})\|$, where $(\hat\beta
_s^o,\hat\beta_s)$ denote the regression coefficients
estimated using only data from study $s$.

%f1
\begin{figure}[b]

\includegraphics{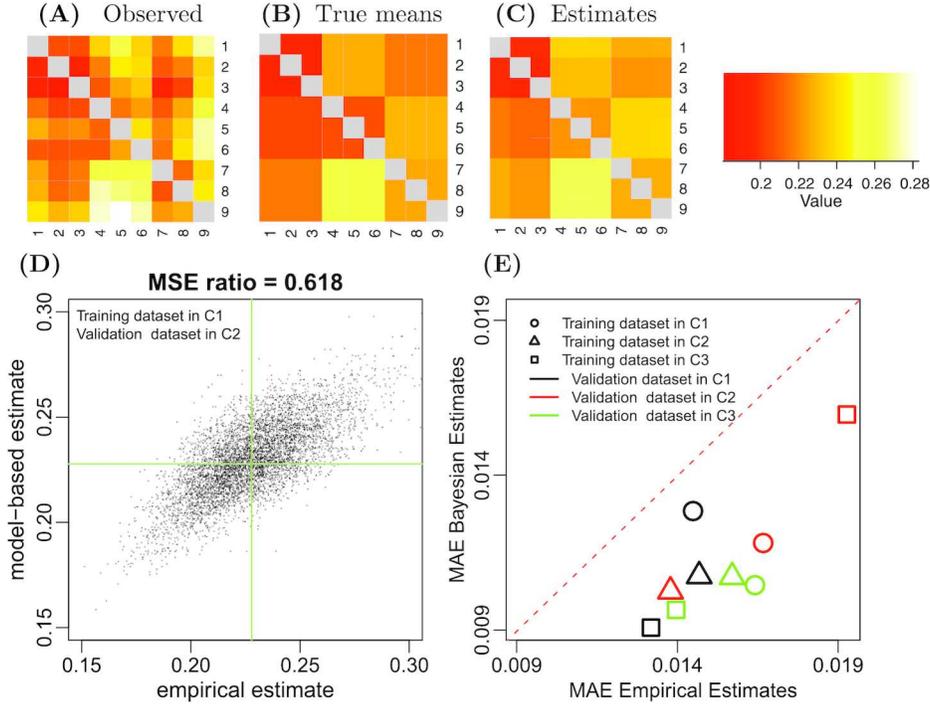}

\caption{Leave-one-in array, with rows corresponding to training data sets and
columns to validation data sets.
Panel \textup{(A)} shows the leave-one-in array $Z$ for a single simulation.
%The diagonal is blank {\tt giusto?}. si aggiunto
Panel \textup{(B)} shows the true expected values $\zeta_{s,v}$ of $Z_{s,v}$. %
%variables obtained integrating with respect to the described sampling
%models $P_1,\ldots,P_S$.
Panel \textup{(C)} shows the Bayesian estimates $\mathbb{E}(\mu_{C(s),C(v)}| Z)$.
The diagonals in panels \textup{(A)}, \textup{(B)} and \textup{(C)} are blank.
Panel \textup{(D)} considers 500 simulations and plots the empirical estimates
$Z_{s,v}$ against the Bayesian estimates $\mathbb{E}(\mu
_{C(s),C(v)}| Z)$. Panel \textup{(D)} considers a training data set $s$ in
$C_1$ and a validation data set $v$ in $ C_2$.
%$s\in C_{j_1}$ and $v\in C_{j_2}$, $j_1,j_2=1,2,3$ .
The green lines correspond to the true expected value $\zeta_{s,v}$.
Panel \textup{(D)} also reports the MSE ratio contrasting the Bayesian estimates
with the empirical estimates.
Panel \textup{(E)} contrasts the Bayesian estimates of $\zeta_{s,v}$ with the
empirical estimates by displaying the MAEs.
Panel \textup{(E)} considers all combinations with $s$ and $v$ in $C_1,C_2$ or $C_3$.}
\label{fig1}
\end{figure}

Figure~\ref{fig1}(A) shows the $Z$ array for a single simulation, with rows
corresponding to training data sets and columns corresponding to
validation data sets.
This array shows that sampling variability accounts for a relevant
part of the observed differences across validation summaries, and the
resulting panel is not easily interpretable by direct visual inspection.
Figure~\ref{fig1}(B) shows Monte Carlo approximations of the
true expected values $\zeta_{s,v}$ of the $Z_{s,v}$ variables under the
described sampling models.
The expected value $\zeta_{s,v}$ is computed integrating with respect
to the actual distributions $(P_s,P_v)$ of
$(X_{s,i},Y_{s,i})$ and $(X_{v,i},Y_{v,i})$.
Figure~\ref{fig1}(C) shows the cluster-based Bayesian estimates $\mathbb{E}(\mu
_{C(s),C(v)} | Z)$ based on our two-stage procedure.
In this simulation, our clustering procedure gives a point estimate
$\hat\Pi=[\{1\},\{2,3\},\{4,5,6\},\{7,8,9\}]$ of the latent partition.
The distance $l(\hat\Pi,\Pi_{\mathrm{TRUE}})$, measured with the maximum
transfer metric, is equal to 1.

%In the illustrated cross-study validation,
Comparison of panels (A) and (C) shows that the two-stage procedure
correctly reconstructs the block structure of the
true expected values $\zeta_{s,v}$ displayed in panel~(B).
Also, the procedure correctly identifies a group of studies, which are
not affected by measurement errors,
with estimated $\mu_{C(s),C(s)}$ value below~0.2.

%GP {\tt volevo riscrivere ma non sono riuscito. A e B sono
%indipendenti dal modello bayesiano. il resto del paragrafo l'ho letto
%tre volte senza capire niente}
% hai ragione: panels A and C ( e non A and B) aggiustato

We repeated the simulation 500 times.
In each iteration, and
for each pair $(s,v)$, we estimated the unknown $\zeta_{s,v}$ means
using our Bayesian estimator $\mathbb{E}(\mu_{C(s),C(v)} | Z)$ and
the empirical estimator $Z_{s,v}$.
The results are plotted in Figure~\ref{fig1}(D) against each other for a single
$(s,v)$ combination, with $s$ in $C_1$ and $v$ in $C_2$.
%Under this simulation scenario, with three groups, $C_1$, $C_2$ and
%$C_3$, we have 9 possible combinations,
%We considered the 9 combinations
%and, for each one,
Then, for each $(s,v)$ combination, we contrasted the MSEs and the MAEs
of the Bayesian estimates with the empirical estimates.
Across all $(s,v)$ combinations the Bayesian estimator has lower MSE
and MAE than the empirical estimates.
These results are graphed in panel (E); each point corresponds to one
$(s,v)$ combination,
and the MAEs of the Bayesian and empirical estimates are plotted
against each other.
In this comparison
the Bayesian estimator achieves a substantially lower dispersion
around the true
expected value $\zeta_{s,v}$ compared to the empirical estimator.

For each simulation we computed $l(\hat\Pi,\Pi_{\mathrm{TRUE}})$, the
number of elementary set operations between the true and estimated
latent partition. On average this distance is 1.63 and, in most
iterations, $\hat\Pi$
has a distance of 2 set operations or less from~$\Pi$.
%%
%For each simulation we also computed posterior confidence regions for
%$\Pi$.
%These are subsets centered at $\hat\Pi$; i.e. they take the
%form $\{\Pi: l(\Pi,\hat\Pi)\le\triangle\}$, where the quantity
%$\triangle$ depends on the desired confidence level.
%When we set the confidence levels at 0.7, 0.8 and 0.9, the resulting
%coverage probabilities are 0.65, 0.77 and 0.88 respectively.

\subsection{Scenario 2}

We consider a sampling model previously used in \citet{waldron2011optimized}.
We use it
to investigate how the comparison of alternative algorithms
is enhanced by Bayesian modeling of the $Z$ arrays.
% as an example of a high-dimensional prediction scenario which favors
%ridge regression over LASSO regression.
Here we add measurement errors to the outcome variable in subsets of studies.
We investigate how modeling of $Z$ allows algorithm performance
assessment for continuously varying training sample size. % within a
%homogeneous dataset cluster. % and clustering based validation
%measures.
The main focus is on the maps $j\rightarrow Z^j_{B(s),s}$ to contrast methods.
We also highlight how
posterior inference on clustering based statistics, such as the
estimates of $\mu_{C(s),C(v)}$, captures uncertainty on the algorithms'
performances.

We simulated 540 zero-mean Normal predictors $X_{s, i}$ with a
covariance matrix structured in blocks:
\[
\sigma_{l,j} =\cases{ %
1, & \quad\mbox{if }$l=j$,
\vspace*{2pt}
\cr
0.2, & \quad\mbox{if }$l,j\le100 \mbox{ and } l\neq j$,
\vspace*{2pt}
\cr
0.2, & \quad\mbox{if }$100< l,j\le 200 \mbox{ and } l\neq j$,
\vspace*{2pt}
\cr
0.2, & \quad\mbox{if }$200< l,j\le370 \mbox{ and } l\neq j$,
\vspace*{2pt}
\cr
0, & \quad\mbox{otherwise}.}
\]
Conditionally on these predictors, we then generated binary outcomes
$Y_{s,i}$ with
$\mathbb{E}(Y_{s,i}| X_{s,i}) =[1 + \exp(-\beta X_{s, i})]^{-1}$.
Here $i\le n_s=100$ and $s=1,\ldots,9$.
%binomial distribution with weights determined by a linear combination
%of predictor variables with inverse-logit link:
The regression coefficients $(\beta_1,\ldots,\beta_{540})$ are $\beta
_{j} = 0.2$ for $j\le370$ and $\beta_j= 0$ for $j>370$.
Departing from the sampling model of \citet{waldron2011optimized}, we
added measurement errors to $Y_{s,i}$, by changing the value of
$Y_{s,i}$ with probability
$0.05$ in $C_1=\{1,2,3\}$, 0.25 in $C_2=\{4,5,6\}$ and $0.5$ in $C_3=\{
7,8,9\}$. These probabilities are independent of $Y_s$ and $X_s$.
Note that any classification approach applied to studies $s=7,8,9$ has
an average error rate of $0.5$ because the binary outcomes
$Y_{s,i}$, after measurement errors, become independent from the
covariates and $\mathbb{E}(Y_{s,i} | X_{s,i})=0.5$.
%GPQ\\ {\tt spiega perche'. a me sembra un po' strano che uno abbia il
%50 percento no matter what. vale per qualunque dataset, o in media
%sulle covariate? che error rate ha l'algoritmo che dice sempre y=1?}
%fatto
%f2
\begin{figure}

\includegraphics{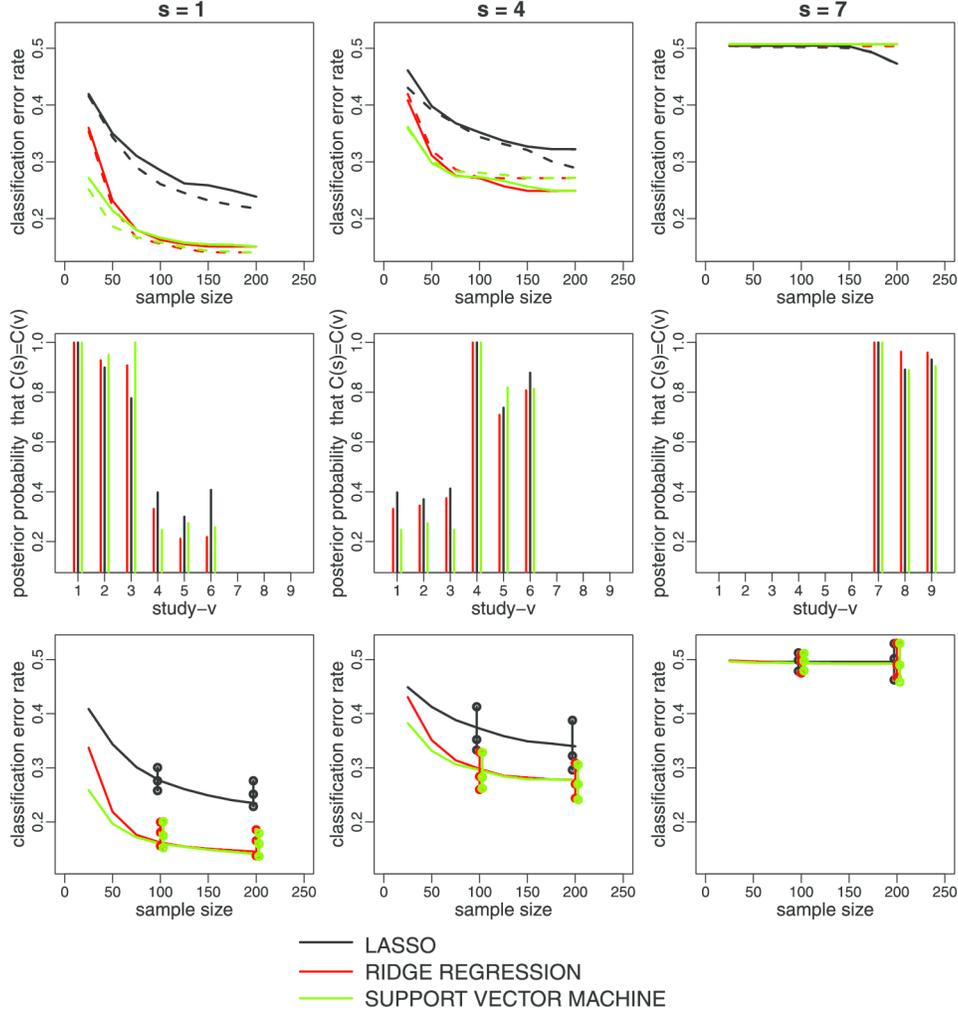}

\caption{Clustering based validation statistics.
The top row considers a single simulation and compares the
%actual {\tt do you mean true or observed?}
true values of the validation statistics
$Z^j_{B(s),s}$ (dashed lines) with our Bayesian\vspace*{1pt} estimates (solid lines)
for varying size of training sets.
Using the same data, the second row displays the posterior
probabilities that two studies, $s$ and $v$, are clustered together.
The third row summarizes results from 500 simulations;
solid lines display the true expected values
$\zeta^j_s$ of $Z^j_{B(s),s}$, while\vspace*{1pt} dots marks medians and quartiles
of the corresponding Bayesian estimates at $j=100$ and $j=200$ across
simulations.
Colors denote the three algorithms.
In the 1st (2nd, 3rd) column $s=1$ $(4,7)$.}\vspace*{-3pt}
\label{fig2}
\end{figure}

We consider for illustration three classification methods:
LASSO regression, ridge regression and a linear support vector
machine; penalization parameters are tuned with cross-validation. We
choose our validation statistics to be the classification error rates.
For each study $s$, we computed the true clustering-based
$Z^j_{B(s),s}$ statistics;
in simulation studies the true latent partition, as well as $B(s)$,
$s=1,\ldots,S$, is known.
% GPQ{\tt true Z?} si true, nelle simulazioni B(s) lo conosciamo, nella
%realta' B(s) va stimato
If, for instance, $s=1$, then $B(s)\setminus s=\{2,3\}$, and
$Z^j_{B(s),s}$ measures the average classification performance
obtained when a model is trained by $j\le200$ records randomly sampled
from $B(s)\setminus s$.
The classification performance is obtained through empirical validation
on data set $s$.

We then used the posterior distribution of $B(s)\setminus s$
and computed the estimates $\mathbb{E}( Z^j_{B(s),s} | Z, \sum_{v\in B(s)\setminus s } n_v \ge j)$.
The first three panels in Figure~\ref{fig2} contrast $Z^j_{B(s),s}$ (dashed
lines) with the Bayesian estimates (solid lines); each color
corresponds to one of the three methods.
Overall, the estimates correctly
%GPQ{\tt qual'e' il gold standard?} ho aggiunto dopo questa frase :
%"These discrepancies are shown in the third row of Figure~2 where the
% maps $j\rightarrow\mathbb{E}(Z^j_{B(s),s})$ are plotted. "
portray the differences that exist between the performances of the
three algorithms; in this scenario,
the support vector machine slightly outperforms ridge regression,
which, in turn, has lower prediction errors than LASSO.
These differences are shown in the third row of Figure~\ref{fig2} where\vspace*{1pt} we plot
the maps $j\rightarrow\zeta^j_s$, with $\zeta^j_s$ equal to the
expected value of $Z^j_{B(s),s}$.
%The estimates also highlight differences between the clustering-based
%validation statistics for studies 1, 4 and 7:
%datasets generated with higher measurement error probabilities tend to
%have higher estimated classification error rates.
The second line of panels in Figure~\ref{fig2} shows the posterior probabilities
$p(C(s)=C(v)| Z)$. In this example, the proposed model captures the
underlying partition of the 9 studies and
the differences across methods' performances.

We repeated the simulation 500 times, generating 9 independent data
sets for each iteration.
In the bottom three panels of Figure~\ref{fig2}, we show medians and quartiles
of the $Z^j_{B(s),s}$ posterior estimates, for $j=100,200$, obtained
across these 500 iterations.
These are compared to approximations of the true maps
$j\rightarrow\zeta^j_s$, obtained by
averaging the true error rates $Z^j_{B(s),s}$ across simulations.
These maps are displayed with solid lines in the third row of panels in
Figure~\ref{fig2}.\vadjust{\goodbreak}
These panels summarize the distribution across simulations of the
estimated clustering validation measures $Z^j_{B(s),s}$
and confirm that the estimates are representative of the performances
of the algorithms being compared.

\section{Application to survival prediction in cancer}\label{app}

We illustrate an application to the development of a
prediction model for overall survival of ovarian cancer patients using
microarray gene expression
data.
%%
%This work is part of a broader effort at our institution to construct
%and validate
%a prediction model, with the final goal of supporting clinical decisions.
%The two major steps of this project will be
%(i) validation of published gene expression signatures using all
%relevant publicly available data, and
%(ii) validation of the best available model(s) using independent
%specimens from a new, appositely designed study.
%
Ovarian cancer is the most lethal gynecological cancer, and numerous
groups have undertaken microarray experiments to measure tumor gene
expression for development of prognostic models of patient survival.
It is widely accepted that gene signatures proposed for clinical
application must be validated on independent data sets. In this area of
research several strategies and methods have been proposed for
prediction. Which one works best? How much uncertainty is involved
in ranking methods? Posterior probabilities on the $\mu_{C(s),C(v)}$
random variables are suitable for answering these questions.

We identified nine previously curated studies
utilizing five different microarray
platforms, each providing patient
overall survival for at least 40 late-stage, serous-type, ovarian
tumors (Table~\ref{tab1}). Microarray data were processed using standard
normalization methods, after which probe identifiers were mapped to
standard gene symbols, as provided by the \emph{curatedOvarianData}
library [\citet{Ganzfried2013}].
Only gene symbols represented on all platforms
were considered for across-platform comparability. We noted that
limiting consideration to those genes present across all platforms has
a negligible effect on prediction performance. For example, we
separately fitted Cox models with ridge penalty and estimated with
cross-validation C-statistics, separately considering one study at a
time; the average decrease of the C-statistics when only genes present
in all platform were considered compared to using all available genes
was less than 0.01.
%t1
\begin{table}
\tabcolsep=0pt
\tablewidth=\textwidth
\caption{The nine ovarian cancer data sets considered in this study.
We only considered late-stage serous tumors from these
studies}\label{tab1}
\begin{tabular*}{\tablewidth}{@{\extracolsep{\fill}}lccc@{}}
\hline
$\bolds{s}$ & \textbf{Study} & $\bolds{n}_{\bolds{s}}$ & \textbf{Microarray platform} %& survival & follow-up & \%
%censoring
\\
\hline
1 & \citet{bentinkangiogenic2012} & 117 & Illumina Human v2 %& 38 & 74
%& 44
\\
2 & \citet{crijns2009} & 157 & Operon Human v3 %& 25 & 24 & 28
\\
3 & \citet{yoshihara2010}& 110 & Affymetrix hgug4112a %& 53 & 40 & 58
\\
4 & \citet{bonome2008} & 185 & Affymetrix hgu133a %& 46 & 100 & 30
\\
5 & \citet{tothillnovel2008} & 139 & Affymetrix hgu133plus2 %& 40 & 40
%& 48
\\
6 &The Cancer Genome Atlas Research Network
(\citeyear{tcga2011})& 420 & Affymetrix hthgu133a %& 41 & 257 & 44
\\
 7 & \citet{mok2009}& \phantom{0}53 & Affymetrix hgu133plus2 %& 25 & 16 & 23
\\
8 & \citet{konstantinopoulos2010} & \phantom{0}42 & Affymetrix hgu95av2 %& 45 & 23
%& 45
\\
9 & \citet{dressman2007}& \phantom{0}59 & Affymetrix hgu133a %& 42 & 23 & 39
\\
\hline
\end{tabular*}
%%Median survival and follow-up times (in months)
% were calculated by the Kaplan-Meier and reverse Kaplan-Meier
%estimate.
\end{table}

\subsection{Accounting for different sample sizes}\label{adj}
The sample size $n_s$ varies across studies.
One can therefore expect higher
values of the validation statistics $Z_{s,v}$ for those models trained
in the largest studies.
To prevent this from creating artifactual clusters of studies, we apply
an intuitive correction.

We selected a threshold of 110 samples and considered the 6 studies
that have a sample size larger than the threshold.
We then computed the empirical
estimates $Z^{110}=(Z^{110}_{s,v}; s,v=1,2,3,4,5,6, s\neq v)$.
%%These estimates reduce the artificially large differences between the
%$Z$'s, by removing variability arising from different sample sizes.
The computation of $Z^{110}_{s,v}$ is straightforward.
We iterate two steps: (i) we train a prediction model $M^{110}_{s}$ with
110 data points sampled without replacement from the $s$th data set,
then (ii) we validate the resulting model
on the entire $v$th data set. %**is the v-th dataset sampled as well?
%does it matter?**
We set $Z^{110}_{s,v}$ equal to the average value
of the validation statistics across iterations. The statistic
$Z^{110}_{s,v}$ estimates
the performance of a model trained by 110 samples from $P_s$. We
computed these estimates with 200 iterations.

The covariance matrix $\Sigma^{110}$ of $ Z^{110}$ is then estimated
by bootstrapping.
%%
%all datasets are resampled with replacement. This produces "artificial"
%datasets with different sample sizes $n_s$
%that are then used for resampling $Z^{110}$. Finally the resampled
%arrays allows one to
%compute the covariance estimate $\hat\Sigma^{110}$.
%
The array $Z^{110}$ and $\hat\Sigma^{110}$
are used for obtaining the posterior distribution of the random
partition $\Pi^{110}$
with the model proposed in Section~\ref{sec2}; we only replace $(Z,\Pi)$
with $(Z^{110},\Pi^{110})$. %We use $p(\Pi^{110}| Z^{110},\hat
The reported probability that two studies, say, $s=1$ and $v=6$, belong\vspace*{1pt}
to the same cluster is provided by
%obtained from
the posterior distribution
$p(\Pi^{110}| Z^{110},\hat\Sigma^{110})$. % on the partition of
%studies.

Next, we need to extend this posterior, which refers to the 6 studies
we selected,
to the remaining 3 which\vspace*{1pt} have less than 110 samples.
To achieve this goal, we compute $p(\Pi^{42}| Z^{42},\hat\Sigma^{42})$ by reducing the
threshold from 110 to 42, and % . This is the distribution of a random
%partition of our 9 studies.
%We
report the following adjusted random partition:
\begin{equation}
\label{pseudpost}
\qquad\hat p(\Pi=\pi)= \frac{ p(\Pi^{42} =
\pi| Z^{42},\hat\Sigma^{42})\times
p(\Pi^{110}=\Delta^{110}(\pi)| Z^{110},\hat\Sigma^{110})}{
\sum_{\pi'}\mathbf{1} ( \Delta^{110}(\pi')=\Delta^{110}(\pi) )
p(\Pi^{42} =\pi' | Z^{42},\hat\Sigma^{42})},
\end{equation}
where the sum is over possible partitions of the 9 studies and the
operator $\Delta^{110}$ projects them into partitions of the 6 studies
$\{1,2,3,4,5,6\}$ above the 110 samples
threshold. Two of these 6 studies $(s,v)$ are clustered together
by $\pi$ if and only if they are clustered together\vspace*{1pt} by $\Delta^{110}(\pi)$.
Expression (\ref{pseudpost})
implies $\hat p(\Delta^{110}(\Pi) =\cdot )=p(\Pi^{110}=\cdot|
Z^{110},\hat\Sigma^{110})$.
%We computed the conditional distributions $p(\Pi^{42}| Z^{42},\hat
%and $p(\Pi^{110}| Z^{110},\hat\Sigma^{110})$ by using the complete
%lists of the possible values for the latent partitions
%$\Pi^{42}$ and $\Pi^{110}$.
%The most computationally intensive stage of the procedure has been the
%computation of $\Sigma^{42}$ and
%$\Sigma^{110}$; the arrays $Z^{42}$ and $Z^{110}$ have been resampled
%1000 times.

This correction for sample size effects preserves the interpretability
of the clustering algorithm.
It also avoids more complex constructions, such as replacing the latent
random variables $\mu$ in (\ref{fst})
with latent functions for sample size-specific
average validation statistics. The most\vspace*{1pt} computationally intensive stage
of the procedure is the computation of $\Sigma^{42}$ and
$\Sigma^{110}$; the arrays $Z^{42}$ and $Z^{110}$ have been resampled
1000 times.

%The outlined approach can be modified by adding thresholds and varying
%their values.
%In our application we observed negligible variations on the reported
%distribution $ \hat p(\Pi=\pi)$ with respect to the choice of
%these thresholds.

\subsection{Comparative analysis of prediction methods}

Ovarian cancer studies for developing prognostic signatures are
commonly based on two distinct groups of data sets, a training group,
which in most cases
only includes a single data set, and a group of publicly available
validation data sets.
%are commonly researchers commonly
%in some single combination of training and validation sets.
A recent example that presents key questions related with our study is
\citet{kang}, and the subsequent comment \citet{kang2}.
The goal in \citet{kang} is to develop a molecular score based on
expression of 151 genes that are involved in platinum-induced DNA
damage repair to predict response to chemotherapy. This exemplifies
using a biological hypothesis to preselect predictors for constructing
prognostic models, thus avoiding some of the challenges arising in the
``large $p$ small $n$'' setting.
%%
%The main
%proposal in \citet{kang} is an innovative approach based on a
%biological hypothesis for constructing prediction models. The
%biological hypothesis is used to select 151 genes that are then used
%for fitting a prediction model using established methods, thus
%avoiding the so called ``large $p$ small $n$'' setting.
%%
In
\citet{kang2} authors point out that both independent validations and a
suitable sample size of the validation data set are essential for
assessing prediction models.

Prescreening the space of predictors has the advantages of parsimony
and interpretability, but comes at the cost of some information loss.
Our goal in this section is to quantify this trade-off using
cross-study validation.
We use the truncated $C_{\tau}$ statistic as proposed in
\citet{unoc-statistics2011}, truncated at $\tau=3$ years, for measuring
survival prediction accuracy.
The $C_{\tau}$ statistic, given a prediction model~$M$ and independent
(possibly censored) survival data with covariates $(Y_i,X_i)$,
$i=1,\ldots,n$, from an unknown
distribution $P$, estimates the conditional probability
$P(r_{n+1}\ge r_{n+2} | Y_{n+1} \le Y_{n+2}, Y_{n+1} \le\tau)$.
The random variables $(r_{n+1}, r_{n+2})$ are risk scores computed
from $M$ based on individual\vspace*{1pt} covariates $(X_{n+1}, X_{n+2})$;
if, for instance, $M$ is a proportional\vspace*{1pt} hazards model with
coefficients $\hat\beta$, then $r_{n+1}=\hat\beta X_{n+1}$ and
$r_{n+2}=\hat\beta X_{n+2}$.
The estimate converges, under the assumption of noninformative censoring,
to the unknown conditional probability.
It is only required that the censoring cumulative distribution
function remains below 1 at $\tau$.

We applied our method with prediction models constructed using several
approaches.
The first one is a direct application of survival ridge regression,
using available gene expression data, under the assumption
of proportional hazards with a linear link function.
The second is similar to the approach followed in \citet{kang}; we only
use available gene expression data
within the selective list proposed by the authors. Note that we do not
attempt to reproduce their study; we follow a similar
strategy. Also, in this case prediction models were derived using
penalized maximum likelihood.
Additionally, to these two approaches we also attempted the use of
Kernel-based methods for estimating a smooth nonparametric link
function [\citet{LiL}].
This produced results (i.e., values of the $C_{\tau}$ estimator)
clearly inferior to the first two approaches.

Our goal is to show that the cross-study validation approach we present
here facilitates
methods comparison by estimating
the easily interpretable $\mu$ latent variables.
All the analyses were repeated separately under
each method.
Modeling of the $Z$ array, in this example, produces an appreciable
reduction of the uncertainty
on the $\mu$ latent variables compared to the direct computation of
the credible intervals by bootstrapping. All the model-based estimates
of the $\mu$ latent variable are shrunk toward
the average of the $Z$ entries.
%f3
\begin{figure}[t]

\includegraphics{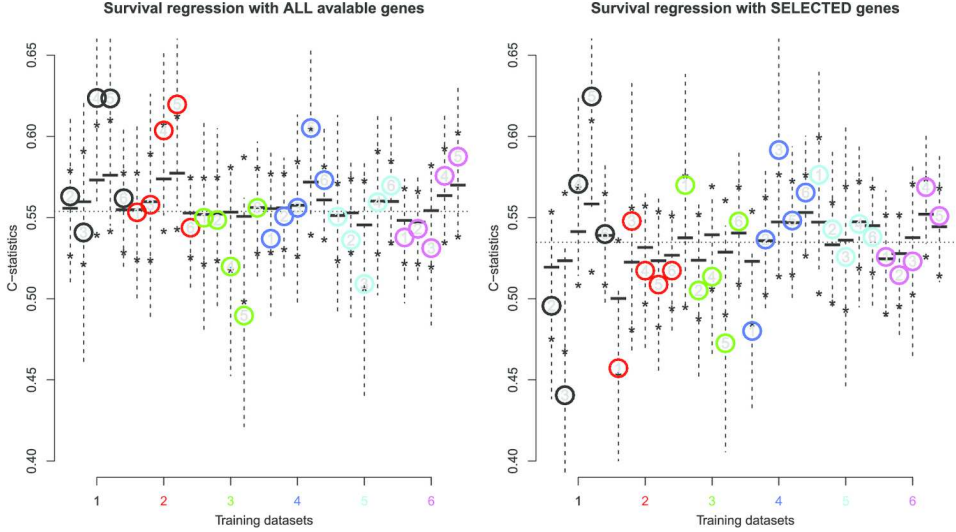}

\caption{Validation analysis based on $C_{\tau}$ statistics. The left
panel considers ridge regression based on all available gene expression
data, while the right panel considers only
a list of genes selected on the basis of the proposal in \citet{kang}.
Each colored point illustrates a $Z^{110}_{s,v}$ validation statistic;
colors indicate the training data set $s$ while
the integers in gray indicate the validation data set $v$.
The ``$-$''
symbols indicate the corresponding Bayesian estimates $\mathbb{E}(\mu^{110}_{C(s),C(v)}| Z^{110})$.
The dashed lines are 80\% confidence intervals of the unknown means
$\mathbb{E}(Z^{110}_{s,v})$ obtained by Bootstrapping.
The ``${}^*$'' symbols indicate 80\% credible intervals obtained from the
posterior distribution of $\mu^{110}_{C(s),C(v)}$ given $Z^{110}$.}\label{fig3}
\vspace*{-6pt}
\end{figure}

Figure~\ref{fig3} shows the observed validation statistics $Z_{s,v}^{110}$.
As mentioned, these are empirical estimates
of predictive performances adjusted\vadjust{\goodbreak} for sample size variability.
Each panel corresponds to one of the two approaches we compare, and
colors indicate the training data sets, while the integers displayed in
grey indicate the validation data sets.
The plots show the 80\% confidence intervals of the unknown means
$\mathbb{E}(Z_{s,v}^{110})$
obtained by bootstrapping (dashed lines). They also
display the model estimates (marked with the ``$-$'' symbol) of the $\mu
_{C(s),C(v)}^{110}$ variables\vspace*{1pt} and
the 80\% credible intervals (marked with the ``${}^*$'' symbol).
Under both approaches all $\mu_{C(s),C(v)}^{110}$ variables
are estimated within the $(0.5,0.6)$ interval.
Our comparison suggests that
models fitted after upfront selection of a subset of genes based on
biological hypothesis perform worse than
using all gene expression data.
This indicates that genes other than those involved directly in DNA
damage repair can contribute to explaining survival of ovarian cancer patients.
All comparisons of the Bayesian estimates $\mathbb{E}(\mu
^{110}_{C(s),C(v)}| Z^{110})$
under the two approaches are consistent with this evaluation.
We also compared the posterior distributions of $\mu_{C(s),C(v)}^{110}$
under the two approaches; at each\vspace*{1pt} pair $(s,v)$, when we sample from the
posterior distributions we obtain inferior $\mu_{C(s),C(v)}^{110}$ values\vspace*{1pt}
for the selective approach with probability above 0.67.
If we use all genes, we obtain a probability of 0.78 that all $\mu
_{C(s),C(v)}^{110}$ are larger than 0.5, meaning that all models
perform better than assigning risk scores completely at random.
%f4
\begin{figure}[b]

\includegraphics{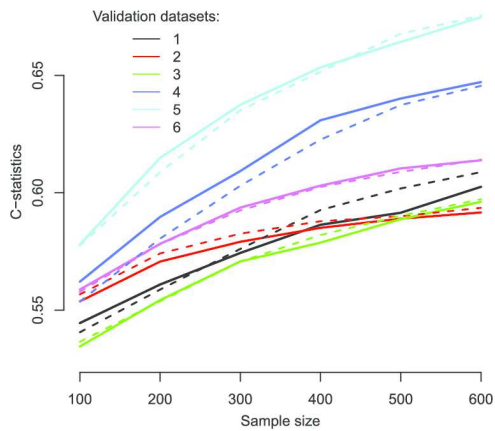}

\caption{Sample size adjusted validation statistics. Solid lines
display estimates of the clustering-based statistics $Z^j_{B(s),s}$
for values of $j$ ranging between 100 and 600. Dashed lines display the
validation statistics $Z^j_{\{1,\ldots,S \},s}$, that is, the cluster $B(s)$
is replaced with the entire collection of 9 studies. Each color
corresponds to a specific validation study $s$.}\label{fig4}
\end{figure}

The posterior distribution of the latent partition of the 6 studies with
sample sizes above 110 suggests the existence of two clusters, one
including studies 4 and 5 and the other with all remaining studies.
The estimate of the latent partition is identical under the
two considered approaches for constructing predictive models but needs
to be combined with relevant uncertainty. In particular, in both
cases the partition constituted by a single degenerate subset with the
six studies together accumulates posterior probabilities from both
approaches above 0.15. When we added to the
analysis the remaining three studies with sample sizes below 110,
the resulting probabilities of the degenerate partition remained above
0.12. In summary, we found moderate evidence of
a nondegenerate partition with heterogeneous subgroups.

In order to interpret the latent partition estimate of our leave-one-in
analysis, we computed clustering-based validation statistics.
Each solid line in Figure~\ref{fig4} is representative of a data set $s$ and
illustrates, for hypothetical sample sizes $j$ from 100 to 600,
estimates of how well outcomes in study $s$ can be predicted by
randomly selecting $j$ data points within the
cluster containing $s$.
More formally, the $y$ axis shows estimates of
%$\mathbb{E}(
$Z^j_{B(s),s}$ %| Z, \sum_{v\in B(s)\setminus s } n_v \ge j)$,
with $j=100,\ldots,600$.
In this example the reported probabilities of the events $ \sum_{v\in B(s)\setminus s } n_v \ge j$ are all above 0.6
for $j\le300$.
These estimates are contrasted (dashed lines in Figure~\ref{fig4}) with
$Z^j_{\{1,\ldots,S \},s}$.
These are estimates of how well outcomes in $s$ can be predicted by
randomly selecting $j$ data points from all available studies.
We observe little difference between the solid and dashed lines.
This similarity suggests that the clustering is not driven by heterogeneous
data quality levels across studies (i.e., there is no evidence of
clusters that produce prediction models with poor performance). Clustering
is driven
by studies 4 and 5, in which, due to covariates distributions, it
appears relatively easier to achieve C-statistics above 0.6 compared to
all other studies.
Clustering-based statistics in Figure~\ref{fig4} suggest additional samples,
above $600$ and above the overall number of samples from the nine
studies, might significantly contribute obtaining better prediction models.

For a comparison, we fitted the data sets with a hierarchical
proportional hazards model, with studies clustered through a Dirichlet
process, and Normal marginal priors for the
regression coefficients.
The prior assigns a vector of regression coefficients to
each cluster of studies, while coefficients are independent across
clusters. To facilitate the comparison, we
tuned the Dirichlet process to match the estimate of the number
of clusters in our leave-one-in analysis.
We used the list of possible values for the latent partition and
approximated the posterior using the approach
discussed in \citet{sinha2003bayesian}.
Our interest is in comparing the latent partitions obtained using the
model just described to those from the validation
analysis.
If clustering in the leave-one-in analysis
is driven by differences in the study-specific regression
models, and not in the distributions of the predictors, then one expects
the two approaches to infer similar partitions.
Instead, the total variation distance between posterior
distributions is 0.79, and we did not notice
similarities. This is consistent with the
interpretation of the
partition inferred through the validation analysis that we
discussed in the previous paragraph.

We also considered \textit{random survival forests} for constructing
prediction models; we used methodology and software discussed in
\citet{ishwaran2008random}. This method directly provides mortality
scores for each sample in a test data set. Under several choices of
the tuning parameters involved in the application of random forests,
including minimal final nodes sizes [see \citet{ishwaran2008random}
for details], the resulting predictive models appeared
inferior to ridge regression when compared using $C_\tau$ statistics.
Under all considered choices of the tuning parameters at least 66\%
of the $\mu^{110}_{C(s),C(v)}$ estimates were inferior to ridge regression.\vspace*{1pt}
Contrasting results with random survival forests based on all
available gene expression data versus the use of selected genes as
suggested in \citet{kang}, we obtained again higher $\mu^{110}_{C(s),C(v)}$
estimates using all available gene expression data.

\section{An example of heterogeneous studies}
Next we discuss cross-study validation of four nonsmall cell lung
cancer studies recently
reviewed in \citet{ferte2013impact}, based on the data sets curated by
the authors.
The data structure is similar to the previous example and includes gene
expression
predictors and patient survival times.
We refer to \citet{ferte2013impact} for a detailed description. The four
studies and corresponding samples sizes are as follows: the Director's
challenge study \citet{shedden2008gene} (299),
\citet{zhu2010prognostic} study (62), \citet{hou2010gene} study (79) and
the TCGA [\citet{hammerman2012comprehensive}] study (90).
This example emphasizes the necessity of accounting for study heterogeneity
and that averaging the $Z_{s,v}$ statistics does not
provide a complete description of models' performances.

The Director's study [\citet{shedden2008gene}] includes data from 4
institutions. Our first analysis
investigates whether there are large differences in the data
originating from these institutions.
The posterior of the model assigns probability 0.83 to the event that
these 4 data sets are clustered together.
Then, we considered the hypothesis of clusters involving the remaining
data sets
[\citet{zhu2010prognostic,hou2010gene,hammerman2012comprehensive}].
The approach is identical to the description in the previous section:
we trained models
using gene expression data and validated using concordance $C_{\tau}$
statistics. The posterior of $\Pi$ produced an estimate of two
clusters, one including
the Director's study and the Zhu et al. study, and the second
including the remaining two studies.
The posterior strongly supports the hypothesis of separate clusters.
The posterior
probability at the estimated configuration $\hat\Pi$ is 0.44, and 0.48
posterior probability accumulates on the neighborhood
$\{\Pi\dvtx  l(\Pi,\hat\Pi)=1\}$.
%f5
\begin{figure}

\includegraphics{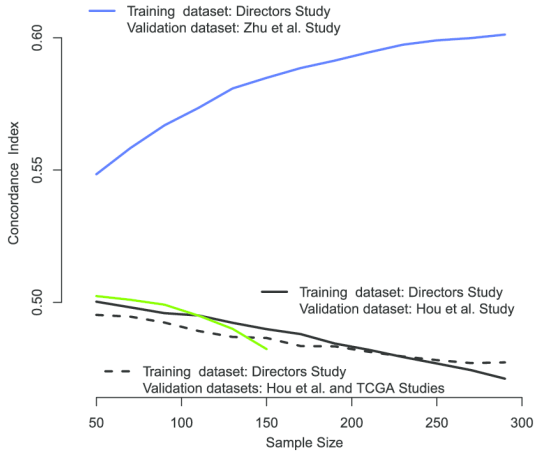}

\caption{Sample size adjusted validation statistics for interpreting
the clustering estimate $\hat\Pi$.
The plot displays $Z_{s,v}^j$ validation statistics when the model is
trained by the largest study [\citet{shedden2008gene}] and validated on
the remaining studies (black and blue lines). Additionally, it displays
validation statistics when we train on the Hou et al. study and use the
TCGA study for validation (green line). Prediction models have been
trained using ridge regression.}\label{lun}\vspace*{-5pt}
\end{figure}

We finally compared sample size adjusted statistics $Z_{s,v}^j$ to interpret
the clustering configuration. Figure~\ref{lun} summarizes the main
discrepancies visualized with
these comparisons. On average, models fitted with $50\le j\le290$ samples
from the Director's study tend to achieve substantially higher
validation results when validated on
the Zhu et al. study (blue line) than when validated on the remaining
two studies (black lines).
In the latter case the validation statistics decrease with
training data set sample size, and %%correct that it is training
%dataset sample size?
the fitted models fail to predict survival times.
We tested this difference using the bootstrap covariance estimates.
The evaluation of prediction models produced by the largest study
[\citet{shedden2008gene}] changes
considerably if we only average the validation statistics across the
three remaining studies, and it appears appropriate
to report substantial discrepancies when we validate the
Director's study results with the Zhu et al.
study, versus validation on the Hou et al. and TCGA studies. While
this is beyond the scope of our analysis, the next step is to
investigate in depth the reason for these discrepancies.

\section{Discussion}
Despite the availability of large collections of related data sets in
many areas of application,\vadjust{\goodbreak} articles that evaluate statistical learning
algorithms based on a comprehensive analysis of available data sets
remain a minority. Those using more than one data set are often based on
cross-validation within each study due to
heterogeneity between studies; see
\citet{japkowicz2011evaluating,demvsar2006statistical} and \citet{bernau2014cross} for
discussions.
%As data availability, model complexity, and societal reliance on
%classification models all grow, it is important to
%establish comprehensive analysis of available datasets as one of the
%standards of algorithm evaluation.
% I am here%
Similar to meta-analyses for
evidence synthesis, comprehensive model evaluations need to jointly
consider study heterogeneity and algorithm performance.
Here we propose
%%
%a general conceptual framework based on the integrated use of
%clustering and model comparison. We implement
%
a Bayesian approach to compare algorithms while incorporating relevant
sources of
uncertainty, including uncertainty on the comparability of independent
studies.

The basis for our framework is the leave-one-in array $Z$ of
validation statistics.
% For a given statistical learning
% algorithm, $Z$ is formed by taking each study in turn, training a
%model
% on that study, and validating it on the remaining independent studies.
%It does not include cross-validation within individual studies.
The concept
is applicable to any validation statistic, such as concordance indices,
classification errors and distances between predicted and observed
responses. While it is certainly possible, and very useful, to simply use
the leave-one-in array as a visualization tool without further modeling,
our experience with evaluating genomic signatures in cancer suggests that
modeling can substantially enhance interpretability of the leave-one-in
analysis. Modeling addresses study heterogeneity, can prevent
erroneous interpretations driven by sampling variability in the summary
statistics, can help address multiplicity issues, and can formalize the
process of identifying outlying studies requiring separate consideration.
The analysis of the $Z$ array helps interpreting the range
of observed cross-study validation statistics, whether it is caused by
differences in the study-specific distributions $P_s$
or it reflects sampling variability.\vadjust{\goodbreak}

Our two-stage procedure is based on a single figure of merit $Z$: this
choice is motivated by the need for a simple strategy and by the
consideration that this still accomplishes the main goal to control
sources of overoptimism such as over-fitting, selection of favorable
training/testing combinations and the use of \mbox{internal} cross-validations
when the studies at hand are heterogeneous. Use of a one-dimensional
figure of merit can, however, be a limitation. For example, if two
studies generated data of poor quality, perhaps because of errors
during sample processing and data management, our algorithm would
likely cluster them together, because they both fail to produce
accurate predictions and generate similarly poor $Z$ scores when used for
validating candidate models. These two studies might still be different
in important ways; for example, they may consider two different
populations. From this perspective, additional summaries of the data
and potentially additional analyses may be advisable to identify
differences between studies.

When multiple studies are available, a natural direction is to combine
them. Bayesian hierarchical models, for instance, have emerged as a
very useful paradigm to borrow information across studies
[\citet{lindley1972bayes} and
\citet{morris1992hierarchical}].
The leave-one-in analysis is not intended to replace combined analyses,
but to address a different question: cross-study replicability of
prediction. We consider the evaluation of prediction methods using
leave-one-in matrices an important complementary goal
and, in some cases, a prerequisite to the
construction of predictive models based on multiple data sources.
The analysis of leave-one-in matrices can be used not only to compare
prediction methods,
but also to select the most appropriate prediction methods for a
subsequent combined analysis.
In a related application to ovarian cancer prognosis using gene
expression profiles, we illustrate a case where we first use
cross-study validation to quantify the extent to which existing
prognostic algorithms can produce results that hold up across studies
[\citet{waldron2014comparative}], and then proceed to develop new
prognostic algorithms based on a combined analysis [\citet{riester2014risk}].

One advantage of the leave-one-in approach is that it can be used to
evaluate any prediction approach, including
heuristic procedures for which it might be challenging to construct
hierarchical extensions. The modeling complexity that comes with
constructing joint models for multiple studies varies across fields. In
some cases the algorithms are based on probabilistic models and
multi-study extensions are possible. In others they are not, and might
be based on heuristics or be very specific to the field of application.
The complexity and problem-specific competence necessary for developing
joint models for heterogeneous data sets are greater compared to the
analysis of $Z$ matrices for off-the-shelf methods.

To address study heterogeneity, we cluster studies with similar
validation profiles through a latent partition. The computation of the
posterior distribution of the latent partition is straightforward and is
a direct application of established\vadjust{\goodbreak} computational strategies for fitting
Dirichlet mixture models. We refer to the supplementary material [\citet{supp}] for
more details.
%%
%The proposed model, and in particular our
%prior distribution on the latent $\mu_{C(s),C(v)}$ variables, is also
%related to recent developments in Bayesian analysis for exchangeable
%arrays \citep{lloydrandom}, whose application has, however, been limited
%so far to relational data. In this context, our proposed prior can be
%described as an exchangeable array, constituted by the $\mu
%_{C(s),C(v)}$ variables, convoluted by a multivariate density $\hat d$
%that captures sampling variability of the $Z$ array.
%%As a side note, our method provides a new clustering strategy for
%%grouping items when
%%one needs to relax the symmetry assumption in the similarity matrix.
%%
%Partitions of studies can arise from different scenarios. For example in
%one scenario, most of the
%studies may have been generated by nearly identical
%sampling models $P_s$, with the exception of a few poorly executed
%studies. Alternatively, distinct populations may have been enrolled in
%various subgroups of studies.
%
Clustering sharpens the interpretation of
the cross-study validation results by allowing
one to explore the maps
$B \rightarrow Z_{B,s}$, focusing on either the estimates $\hat B(s)$
or on those partitions
that a posteriori appear consistent with the dispersion estimate
$\hat{d}$ and the observed array~$Z$.

%Integrating clustering into cross-study validation balances two
%competing needs: on the one hand it is important to perform cross-study
%validation across experiments that are reasonably comparable to each
%other. Comparability should be part of the selection criteria for the
%studies in the first place, but it is also important to make provisions
%for unknown artifacts or biases that can render the cross-study
%comparison problematic. On the other hand it is important to perform
%analyses that are free from the common optimistic bias arising from
%manually selecting combinations of training and validation studies that
%favor an investigator's method or point of view.

%
A simple alternative to formal Bayesian
clustering of data sets is a reordering of rows and columns of $Z$, by
maximizing
objective functions, to obtain high values of the validation statistics
close to the matrix diagonal. While this is perhaps simpler
than what we propose, it can be dangerous to interpret the $Z_{s,v}$
validation summaries without consideration of the associated sampling
variability, and it is easy to introduce an optimistic bias with
clusters obtained by optimizing intra-cluster validation statistics.
%%
%Our approach averages over clusters
%in proportion to the support they receive from the data, and accounts
%for the sampling covariance matrix of the $Z$'s. This protects the
%interpretation from optimistic bias.

In this article we only considered external validation statistics, where
training and testing are performed on separate studies.
%This is
%consistent with our overall goal of providing tools for analyses based
%on a comprehensive look at multiple studies.
Alternatively,
one could integrate internal cross-validation into our framework by
adding a diagonal to the $Z$ array, with entries consisting of within
study cross-validation
statistics. A drawback of standard cross-validation techniques in this
context is that they may result in overly optimistic assessments
[\citet{bernau2014cross}].
%In summary, cross-validation may favor algorithms that adapt
%well to study-specific technological differences, rather than
%algorithms that
%best model the processes of common interest between independent
%studies.
%We think that adding a diagonal can be useful mostly when all
%study specific distributions are identical, i.e. $P_s=P_v$, $s,v\le S$.

%We hope our approach will provide the basis for further work in this
%area.
In this work we compared learning algorithms by
separate analyses of the resulting $Z$ arrays, but
a natural
extension is the joint analysis of multiple $Z$ arrays
corresponding to competing algorithms.
%Also, it is possible to gain additional insight into the
%study partition by borrowing strength across algorithms.
A similar discussion applies to
consideration of multiple validation statistics at the same time. A
separate refinement could seek a data-driven approach for selecting the
thresholds described in Section~\ref{adj} to correct for
sample size differences across studies.

\begin{supplement}[id=suppA]
\stitle{Supplement to ``Bayesian nonparametric cross-study validation of prediction methods''}
\slink[doi]{10.1214/14-AOAS798SUPP} %[doi,text={...}] - jei reikiasuskaldyti doi
\sdatatype{.pdf}
\sfilename{aoas798\_supp.pdf}
\sdescription{We discuss results for
logistic regression, Poisson regression, proportional
hazards models and support vector machine procedures in the
supplementary material.}
\end{supplement}

\printaddresses
\end{document}